\documentclass[a4paper]{article}

\usepackage{amsmath}
\usepackage{amssymb}
\usepackage[arrow,matrix]{xy}

\newcommand{\hoP}{\ensuremath{\mathbb{P}}}

\newcommand{\hoQ}{\ensuremath{\mathbb{Q}}}

\newcommand{\Boxe}{\raisebox{.8ex}{\framebox}}

\newcommand{\C}{\ensuremath{\mathcal{C}}}

\newcommand{\oB}{\ensuremath{\mathbb{B}}}

\DeclareMathOperator{\Coalg}{Coalg}

\DeclareMathOperator{\id}{id}

\DeclareMathOperator{\Hom}{Hom}

\DeclareMathOperator{\Ker}{Ker}

\makeatletter
\renewcommand{\section}{\@startsection
	{section}%				% the name
	{1}%					% the level
	{0mm}%					% the indent
	{-\baselineskip}%			% the beforeskip
	{0.5\baselineskip}%			% the afterskip
	{\bfseries\large}}%			% the style
\makeatother

\makeatletter
\renewcommand{\subsection}{\@startsection
	{subsection}%				% the name
	{2}%					% the level
	{0mm}%					% the indent
	{\baselineskip}%			% the beforeskip
	{-\fontdimen2\font%
	  plus -\fontdimen3\font
	  minus -\fontdimen4\font}%		% the afterskip
	{\bfseries\normalsize}}%		% the style
\makeatother

\newcounter{imlist}
\newenvironment{IMlist}%
   {    \begin{list}{(\roman{imlist})\ }{\usecounter{imlist}%
	\setlength{\labelsep}{0cm} \setlength{\leftmargin}{0cm}%
	\setlength{\labelwidth}{0cm} \setlength{\listparindent}{0cm}%
	\setlength{\topsep}{0cm} \setlength{\parskip}{0cm}%
	\setlength{\partopsep}{0cm}%
	\setlength{\parsep}{0cm}%
	\setlength{\itemsep}{0cm}}}%
   {\end{list}}

\newcommand{\lemma}[1]{ \subsection{Lemma.} \textit{#1} }

\newlength{\afterlemmaskip}
\newcommand{\theorem}[1]{ \subsection{Theorem.} \textit{#1} }

\newcommand{\proposition}[1]{ \subsection{Proposition.} \textit{#1} }

\newenvironment{proof}%
   { \noindent \textit{Proof:\ }}%
   {\hfill $\Boxe$ \vspace{\baselineskip} }

\begin{document}

\mbox{ }
\vspace{0.5cm}
\begin{center}
{\Large On the Connes-Kreimer construction of Hopf Algebras}

\vspace{0.5cm}

{\large I. Moerdijk}
\end{center}

\vspace{0.5cm}

\noindent \emph{Abstract:} We give a universal construction of families
of Hopf $\hoP$-algebras for any Hopf operad $\hoP$.  As a special case,
we recover the Connes-Kreimer Hopf algebra of rooted trees.

\vspace{0.3cm}


\vspace{0.3cm}

\noindent In \cite{K}, \cite{CK} a Hopf algebra $H$ of rooted trees is
discussed.  This algebra originates in problems of renormalisation \cite{K}
and is closely related to the Hopf algebra introduced in \cite{CM} in the
context of cyclic homology and foliations.  The algebra $H$ is the
polynomial algebra on countably many indeterminates $T$, one for each
finite rooted tree $T$.  Its comultiplication is given by the formula
\begin{equation*}
	\Delta(T) = 1 \otimes T + T \otimes 1 + \sum_{c} F_c \otimes R_c,
\end{equation*}
see \cite{CK}.  Here $c$ ranges over all ``cuts'' (prunings) of the tree
$T$.  Such cuts are assumed non-empty, and to contain at most one edge on
each branch.  $R_c$ is the part of the tree which remains after having
performed the pruning, and $F_c$ is the product of subtrees which have
fallen on the ground.  In \cite{CK} it is proved that this 
comultiplication indeed makes $H$ into a Hopf algebra.
% Hopf algebra
Furthermore, $H$
is equipped with a linear endomorphism $\lambda$, which is a universal
cocycle for a suitably defined Hochschild cohomology of Hopf algebras.

The first aim of this note is to show that all these properties can in fact
be deduced from a more basic universal property of $H$.  Namely, $H$ is the
initial object in the category of (commutative unitary) algebras equipped
with a linear endomorphism.  Having realized that this is the case, it
becomes clear that $H$ is in fact equipped with a large family of Hopf
algebra structures, all making the endomorphism $\lambda$ into a universal
cocycle for the corresponding Hochschild cohomology.  For example, for any
two complex numbers $q_1$ and $q_2$, there is a coproduct on $H$, uniquely
determined by the identity
\begin{equation*}
	\Delta(\lambda(T)) = \sum q_1^{|T_{(1)}|} \cdot T_{(1)} \otimes
	\lambda(T_{(2)}) + \lambda(T_{(1)}) \otimes q_2^{|T_{(2)}|} \cdot T_{(2)},
\end{equation*}
where $|T|$ denotes the number of nodes in the tree $T$.  For $q_1 = 1$ and
$q_2 = 0$ one recovers the Hopf algebra structure of \cite{CK}.

The second aim is to describe how this construction applies more generally
to ``algebras'' for any operad $\hoP$ on an additive category, as soon as
one has a well-behaved tensor product of algebras.  More precisely, we will
show that if $\hoP$ is a ``Hopf operad'' on a symmetric monoidal additive
category, then the initial object in the category of $\hoP$-algebras
equipped with a ``linear'' endomorphism is naturally equipped with a family
of natural Hopf $\hoP$-algebra structures.  The algebra of rooted trees
then becomes the extreme instance of this construction where the operad
$\hoP$ is the unit object in each degree.

\subsection*{Acknowledgements.}

My attention was first drawn to the algebra $H$ by A. Connes at the
``Karoubi Fest'' in Paris (November 1998).  I would like to thank Ezra
Getzler and Andr\'e Joyal for helpful discussion.  I am indebted to the
Dutch Science Foundation (NWO) for financial support.  The main results of
this paper were first presented at the Newton Institute, in February 1999.

\section{Operads and algebras.}

\subsection{The underlying category.}
\label{ss1.1}

In this preliminary section we will consider operads on a category $\C$.
We will assume that $\C$ is a symmetric monoidal additive category, with
countable sums and quotients of actions by finite groups on objects
of $\C$.  (In most cases, $\C$ will be closed under all small colimits.)
As an example, the reader may wish to keep the category of vector spaces
over a field $k$ in mind in what follows.  We will write $k$ for the unit
object of $\C$, and $a,l,r$ for the associativity and unit isomorphisms.
The symmetry will be denoted by $c$, with components $c_{X,Y} : X \otimes
Y \rightarrow Y \otimes X$.  We will assume that $\otimes$ is an additive
functor in each variable separately.  Often, the isomorphisms $a, l,
r$ will be suppressed from the notation, and we identify $k \otimes X$
with $X$, and $ X \otimes ( Y \otimes Z)$ with $(X \otimes Y) \otimes
Z$, etc.  This is justified, on the basis of Mac Lane's coherence theorem.
See \cite{CWM} for details.

\subsection{Operads.}

(\cite{M}, \cite{KM}, \cite{GK}, \ldots)  We will consider operads $\hoP$
on such a category $\C$, and write $\hoP(n)$ for the object (of $\C$) of
$n$-ary operations.  We will always assume that our operads have a
distinguished ``unit element'' $u: k \rightarrow \hoP(0)$.  We will
\emph{not} assume that this map is an isomorphism, i.e. that $\hoP$ is
unitary in the sense of \cite{KM}.  Many operads are unitary,
but the constructions of 1.3 lead us out of unitary operads.  Note
that the unit $u: k \rightarrow \hoP(0)$ provides us with a unit $u_A: k
\rightarrow A$ in any $\hoP$-algebra $A$.

The functor underlying the monad on $\C$ whose algebras are $\hoP$-algebras
will be denoted by $F_{\hoP} : \C \rightarrow \C$; so for any object $V$ in
$\C$,
\begin{equation*}
	F_{\hoP}(V) = \coprod_{n \geq 0} \hoP(n) \otimes_{\Sigma_n}
	V^{\otimes n}.
\end{equation*}
This object $F_{\hoP}(V)$ is the free $\hoP$-algebra generated by $V$.

% terrible LaTeX, but the result is rather nice.
\stepcounter{subsection}

\vspace{\baselineskip}

\begin{IMlist}
\setcounter{imlist}{1}
\renewcommand{\theimlist}{\roman{imlist}}
\item[\textbf{\arabic{section}.\arabic{subsection} \ \ Two constructions.} (\theimlist) ]
\label{ss1.3.i}
If $\hoP$ is an operad on $\C$ and $G$ is an object of $\C$, there is
an operad $\hoP_G$ whose algebras are $\hoP$-algebras equipped with a map
from $G$.  Thus, $\hoP_G$ is obtained from $\hoP$ by adding $G$ to the
space $\hoP(0)$ of "constants" (nullary operations).  Explicitly,
\begin{equation*}
	\hoP_{G}(n) = \coprod_{p \geq 0} \hoP(n+p) \otimes_{\Sigma_p}
	G^{\otimes p}.
\end{equation*}
Note that the initial $\hoP_G$-algebra $\hoP_G(0)$ is the free
$\hoP$-algebra $F_{\hoP}(G)$ on $G$.

\stepcounter{imlist}

\item[\hspace{0.45cm} (\theimlist) ] Let $\hoP$ be an operad on $\C$.
A \emph{$\hoP[t]$-algebra} is a pair $(A, \alpha)$ where $A$ is
a $\hoP$-algebra and $\alpha: A \rightarrow A$ is a map in $\C$.
(We will often refer to maps in $\C$ as ``\emph{linear maps}'',
to contrast them with $\hoP$-algebra homomorphisms.)  A map between
$\hoP[t]$-algebras $(A,\alpha) \rightarrow (B,\beta)$ is a map of
$\hoP$-algebras $f: A \rightarrow B$ such that $\beta f = f \alpha$.
This defines a category of $\hoP[t]$-algebras.  This category is the
category of algebras for an operad, again denoted $\hoP[t]$.  It is the
operad obtained by freely adjoining a unary operation ``$t$'' to $\hoP$.
It is not difficult to give an explicit description of $\hoP[t]$ in
terms of trees, analogous to constructions in \cite{GK}.  We will not
need such an explicit description.
\end{IMlist}

\subsection{Example.}

Let $\C$ be the category of vector spaces over a field $k$, and let $\hoP$
be the operad $\hoP(n) = k$.  Its algebras are commutative unitary
$k$-algebras, and the monad $F_{\hoP}$ associated to $\hoP$ is the
symmetric algebra functor.  The associated operad $\hoP[t]$ can be
described as follows.  The space $\hoP[t](n)$ is the vector space on rooted
finite trees T, with one ``\emph{output node}'', the root, and $n$
``\emph{input nodes}'',
labelled by $x_1, \ldots, x_n$.  The \emph{inner nodes} represent application of
the new unary operation $t$.  For example, the tree
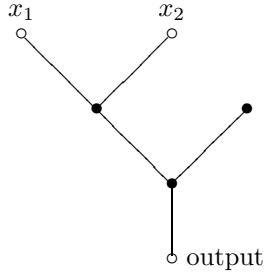
\[
\begin{xy}
% points
<0cm,3cm>="LU"*{\circ}+U*+!D\txt{$x_1$} , <2cm,3cm>="RU"*{\circ}+U*+!D\txt{$x_2$},
         <1cm,2cm>="LUM"*{\bullet} ,            <3cm,2cm>="RUM"*{\bullet} ,
                     <2cm,1cm>="C"*{\bullet},
                     <2cm,0cm>="CD"*{\circ}+R*+!L\txt{output},
% lines
\ar@{-} "LU"+<0.5mm,-0.5mm> ; "LUM",      \ar@{-} "RU"-<0.5mm,0.5mm> ; "LUM",
               \ar@{-} "LUM" ; "C",                \ar@{-} "RUM" ; "C",
                           \ar@{-} "C" ; "CD"+<0mm,0.5mm>

\end{xy}
\]
represents the binary operation $t(t(x_1 \cdot x_2) \cdot t(1))$.  The tree
$\circ$ consisting of just the output vertex represents the element
(nullary operation) $1$.  We will refer to the algebra $\hoP[t](0)$ as the
algebra of \emph{finite rooted trees}.  It can be identified with the
Connes-Kreimer algebra $H$ mentioned in the introduction.  (There is a
slight difference in notation, in that we have merged a product of trees
into one tree with a new output node added to it.)

\section{Hopf operads.}

\subsection{Coalgebras.}

Let $\C$ be a category as in \ref{ss1.1}.  A \emph{coalgebra}
$\underline{X} = (X, \varepsilon, \Delta)$ is an object $X$ of $\C$
equipped with a coassociative comultiplication $\Delta : X \rightarrow X
\otimes X$, and a counit $\varepsilon: X \rightarrow k$ for this
comultiplication.  The associated category $\Coalg(\C)$ is again a
(symmetric) monoidal category, with the usual tensor product
($\underline{X} \otimes \underline{Y}$ is $X \otimes Y$ with as
comultiplication the composition of $\Delta_X \otimes \Delta_Y : X \otimes
Y \rightarrow (X \otimes X) \otimes (Y \otimes Y)$ and the symmetry $X \otimes
c \otimes Y: (X \otimes X) \otimes (Y \otimes Y) \rightarrow (X \otimes Y)
\otimes ( X \otimes Y)$).

\subsection{Hopf operads.}

A \emph{Hopf operad} on $\C$ is an operad $\hoP$ on $\C$ equipped with
additional structure making it an operad on $\Coalg(\C)$.  Thus, each
$\hoP(n)$ has the structure of a coalgebra,
\begin{equation}
\label{eq1}
	k \stackrel{\varepsilon}{\longleftarrow} \hoP(n)
	\stackrel{\Delta}{\longrightarrow} \hoP(n) \otimes \hoP(n),
\end{equation}
this structure is $\Sigma_n$-invariant, and the structure maps of the
operad $\hoP(n) \otimes \hoP(k_1) \otimes \cdots \otimes \hoP(k_n)
\rightarrow \hoP(k_1 + \cdots + k_n)$ are coalgebra maps.  The notion of a
Hopf operad has been introduced in \cite{GJ}. (But beware that their
coalgebras are not necessarily counital.)  I will sometimes write
$\underline{\hoP}$ for this operad on $\Coalg(\C)$, as opposed to the
operad $\hoP$ on $\C$.  The Hopf operad $\hoP$ is \emph{cocommutative} if
each of the coalgebras $\hoP(n)$ is.

If $\hoP$ is a Hopf operad, then the tensor product $A \otimes B$ of two
$\hoP$-algebras $A$ and $B$ is again a $\hoP$-algebra, by the maps
\begin{gather*}
	\hoP(n) \otimes (A \otimes B)^{\otimes n} \stackrel{\Delta \otimes
	\id}{\longrightarrow} \hoP(n) \otimes \hoP(n) \otimes (A \otimes
		B)^{\otimes n} \stackrel{c}{\longrightarrow} \\
	(\hoP(n) \otimes A^{\otimes n}) \otimes (\hoP(n) \otimes B^{\otimes
	n}) \longrightarrow A \otimes B.
\end{gather*}
Moreover, the counits $\varepsilon: \hoP(n) \rightarrow k$ in \eqref{eq1}
make $k$ into a $\hoP$-algebra, which is a unit for this tensor product of
$k$-algebras.  Thus, the category of $\hoP$-algebras is again a monoidal
category (symmetric if $\hoP$ is cocommutative).  A coalgebra in this
category of $\hoP$-algebras is the same thing as a
$\underline{\hoP}$-algebra in the category $\Coalg(\C)$ of coalgebras, and
(as in \cite{GJ}) will be referred to as a \emph{Hopf $\hoP$-algebra}.

\subsection{Example.}

The free $\hoP$-algebra $F_{\hoP}(G)$ on an object $G$ has a canonical Hopf
$\hoP$-algebra structure, cocommutative if $\hoP$ is.  Indeed, since
$F_{\hoP}(G)$ is \emph{free}, the maps $0: G \rightarrow k$ and $\id
\otimes 1 + 1 \otimes \id: G \rightarrow F_{\hoP}(G) \otimes F_{\hoP}(G)$
into $\hoP$-algebras extend uniquely to $\hoP$-algebra maps
\begin{equation*}
	k \stackrel{\varepsilon}{\longleftarrow} F_{\hoP}(G)
	\stackrel{\Delta}{\longrightarrow} F_{\hoP}(G) \otimes F_{\hoP}(G),
\end{equation*}
and one easily checks that this provides the claimed structure.

\section{The Connes-Kreimer construction.}

Let $\hoP$ be a Hopf operad on a category $\C$ as before, and let $\hoP[t]$
be the associated operad whose algebras are $\hoP$-algebras equipped with a
``linear'' endomorphism.  We now present a general construction of Hopf
$\hoP$-algebras, of which the Connes-Kreimer Hopf algebra is a special
case.

\subsection{The initial $\hoP[t]$-algebra.}

Let $(H, \lambda)$ denote the initial $\hoP[t]$-algebra, i.e. $(H, \lambda)
= \hoP[t](0)$.  Thus $H$ is a $\hoP$-algebra, $\lambda: H \rightarrow H$ is
a linear map (i.e. just an arrow in $\C$), and these have the following
universal property:  For any $\hoP$-algebra $A$ and any linear map $\alpha:
A \rightarrow A$, there is a \emph{unique} $\hoP$-algebra map $\varphi: H
\rightarrow A$ such that $\alpha \varphi = \varphi \lambda$.

\lemma{
There is a unique augmentation $\varepsilon: H \rightarrow k$ with $\lambda
\varepsilon = 0$.
}

\vspace{\afterlemmaskip}

\begin{proof}
Apply the universal property to the $\hoP$-algebra $k$ with the zero
endomorphism.
\end{proof}

Next, let $\sigma_1, \sigma_2: H \rightarrow H$ be two linear maps.  Let
\begin{equation*}
	(\sigma_1, \sigma_2) = \sigma_1 \otimes \lambda + \lambda \otimes
	\sigma_2 : H \otimes H \rightarrow H \otimes H.
\end{equation*}
This gives $H \otimes H$ the structure of a $\hoP[t]$-algebra.  So there is
a unique $\hoP$-algebra map
\begin{equation*}
	\Delta = \Delta_{\sigma_1, \sigma_2} : H \rightarrow H \otimes H
\end{equation*}
such that $(\sigma_1, \sigma_2) \circ \Delta = \Delta \circ \lambda$.

% terrible tex, but the result is nice

\stepcounter{subsection}

\vspace{\baselineskip}
\begin{IMlist}
\itshape
\setcounter{imlist}{1}
\label{ss3.3}
\renewcommand{\theimlist}{\roman{imlist}}
\item[\normalfont{\textbf{\arabic{section}.\arabic{subsection} \ \ Lemma.}}
(\theimlist) ]
If $\varepsilon \sigma_i = \varepsilon$ for $i = 1,2$ then
$\varepsilon: H \rightarrow k$ is a counit for $\Delta$.

\item If, in addition, $\Delta \sigma_i = (\sigma_i \otimes \sigma_i)
\Delta$ for $i = 1,2$ then $\Delta$ is coassociative.
\end{IMlist}

\vspace{\afterlemmaskip}

\begin{IMlist}
\setcounter{imlist}{1}
\renewcommand{\theimlist}{\roman{imlist}}
\item[\textit{Proof:} (\theimlist) ] Consider the maps
\begin{center}
\mbox{
\xymatrix{ {(H,\lambda)} \ar[r]^-{\Delta} & {(H \otimes H,
(\sigma_1, \sigma_2))} \ar@<1mm>[r]^-{\id \otimes \varepsilon}
\ar@<-1mm>[r]_-{\varepsilon \otimes \id} & {(H,\lambda),}
}
}
\end{center}
where on the right the isomorphisms $H \otimes k = H = k
\otimes H$ have been suppressed.  By initiality of $H$, it is enough to
prove that $\id \otimes
\varepsilon$ and $\varepsilon \otimes \id$ are $\hoP[t]$-homomorphisms.
This is indeed the case, since
\begin{eqnarray*}
(\id \otimes \varepsilon) ( \sigma_1, \sigma_2) & = & (\id \otimes
\varepsilon)( \sigma_1 \otimes \lambda + \lambda \otimes \sigma_2) \qquad
\text{(definition)} \\
	& = & \sigma_1 \otimes \varepsilon \lambda + \lambda \otimes
\varepsilon \sigma_2 \\
	& = & \lambda \otimes \varepsilon \sigma_2 \qquad (\varepsilon
\lambda = 0) \\
	& = & \lambda \otimes \varepsilon \qquad \text{(assumption)} \\
	& = & \lambda \circ ( \id \otimes \varepsilon),
\end{eqnarray*}
and similarly $(\varepsilon \otimes \id ) ( \sigma_1 , \sigma_2 ) = \lambda
\circ ( \varepsilon \otimes \id ) $.

\stepcounter{imlist}
\item[\hspace{0.45cm} (\theimlist) ] Consider the map $\nu : H \otimes
H \otimes H \rightarrow H \otimes H \otimes H$,
\begin{equation*}
	\nu = \lambda \otimes \sigma_2 \otimes \sigma_2 + \sigma_1 \otimes
\lambda \otimes \sigma_2 + \sigma_1 \otimes \sigma_1 \otimes \lambda.
\end{equation*}
This makes $H^{\otimes 3}$ into a $\hoP[t]$-algebra, so there is a unique
$\hoP[t]$-homomorphism $(H, \lambda) \rightarrow (H^{\otimes 3}, \nu)$.  It
thus suffices to show that $(\id \otimes \Delta) \Delta$ and $(\Delta
\otimes \id) \Delta$ both are.  For the first,
\begin{eqnarray*}
(\id \otimes \Delta) \Delta \lambda & = & (\id \otimes \Delta) ( \sigma_1
\otimes \lambda + \lambda \otimes \sigma_2) \Delta \\
	& = & (\sigma_1 \otimes \Delta \lambda + \lambda \otimes \Delta
\sigma_2) \Delta\\
	& = & (\sigma_1 \otimes \sigma_1 \otimes \lambda + \sigma_1 \otimes
\lambda \otimes \sigma_2  + \lambda \otimes \sigma_2 \otimes \sigma_2)(\id \otimes \Delta) \Delta \\
	& = & \nu ( \id \otimes \Delta) \Delta.
\end{eqnarray*}
The calculation for $(\Delta \otimes \id) \Delta$ is similar. \hfill $\Boxe$
\end{IMlist}

\vspace{\afterlemmaskip}

The preceding lemmas prove:

\theorem{
The initial $\hoP[t]$-algebra $(H, \lambda)$ has a natural family of Hopf
$\hoP$-algebra structures, parametrized by pairs $\sigma_1, \sigma_2: H
\rightarrow H$ satisfying the conditions of Lemma 3.3.
}

\subsection{Example.}

The conditions of Lemma 3.3 are always satisfied if one takes
$\sigma_i$ to be the identity $H \rightarrow H$ or the composition of the
counit $\varepsilon: H \rightarrow k$ and the unit $u: k \rightarrow H$, or
any convex combination $\alpha \cdot \id + \beta \cdot u \varepsilon : H
\rightarrow H$ of these two (for $\alpha, \beta : k \rightarrow k$ with
$\alpha + \beta = \id$).
This provides many different Hopf $\hoP$-algebra structures on $H$.

\subsection{Example.}

Consider again the case of the commutative unitary algebra operad of
1.4.  Then $H$ is the algebra of finite rooted trees $T$.  Note
that $\varepsilon(T) = 0$ as soon as $T$ has at least one inner node.
Write $|T|$ for the number of inner nodes of $T$.  Now let $q_1, q_2 \in k$
be any two numbers, and let
\begin{equation*}
	\sigma_i = q_i^{|T|} \cdot T, \quad \text{for} \ i = 1,2
\end{equation*}
Then $\sigma_1$ and $\sigma_2$ satisfy the condition of Lemma 3.3.
Thus for any two $q_1, q_2 \in k$, the algebra $H$ has a Hopf algebra
structure, with the usual counit, and with comultiplication completely
determined by the identity
\begin{equation*}
	\Delta \lambda(T) = \sum q_1^{|T_{(1)}|} T_{(1)} \otimes
\lambda(T_{(2)}) + \lambda(T_{(1)}) \otimes q_2^{|T_{(2)}|} \cdot T_{(2)}
\end{equation*}
where we write $\Delta(T) = \sum T_{(1)} \otimes T_{(2)}$ as usual
\cite{S}.  For the values $q_1 = 1$ and $q_2 = 0$ one finds $\sigma_1 =
\id$ and $\sigma_2 = \varepsilon$, and one recovers the Hopf algebra
structure of \cite{CK}.

\subsection{Remark.}

The results and examples in this section have been stated for the initial
$\hoP[t]$-algebra $(H, \lambda) = \hoP[t](0)$.  Similar facts hold for the
free $\hoP[t]$-algebra generated by any object $G$ of $\C$.  Writing
$(H[G], \lambda)$ for this algebra and $j: G \rightarrow H[G]$ for the
universal map from $G$, one defines $\Delta: H[G] \rightarrow H[G] \otimes
H[G]$ from $\sigma_1$ and $\sigma_2$ as the unique map of
$\hoP[t]$-algebras satisfying $\Delta \lambda = ( \sigma_1 \otimes \lambda
+ \lambda \otimes \sigma_2) \Delta$ as before and extending the map $u
\otimes j + j \otimes u: G \rightarrow H[G] \otimes H[G]$ (where $u: k
\rightarrow H[G]$ is the unit).  However, rather than doing the calculation
again, this can be seen as a formal consequence of the statements made for
the initial algebra, because the free $\hoP[t]$-algebra on $G$ is the
initial $\hoP_{G}[t]$-algebra (cf. 1.3.(i)), and $\hoP_G$ is a Hopf
operad whenever $\hoP$ is.

\section{Hochschild cohomology.}

In \cite{CK} it is proved that for the Connes-Kreimer algebra $(H,
\lambda)$ (cf. Example 3.6), the map $\lambda$ is a universal
$1$-cocycle for Hochschild cohomology.  In this section, we show that this
result extends to our more general construction.

Recall the definition of the Hochschild cohomology groups $H^*(A,M)$ for
any algebra $A$ and any bimodule $M$, from the complex with maps
$A^{\otimes n} \rightarrow M$ as cochains (see e.g. \cite[formula
(1.5.1.1)]{L}).  Turning around all the arrows in a diagrammatic form
of this definition, one obtains a cohomology $H^*(E,C)$ of a coalgebra $C$
with coefficients in a bicomodule $E$, as the cohomology of the complex
$C^n(E,C) = \Hom_{\C}(E, C^{\otimes n})$.  Explicitly, this is the
cohomology of the simplicial abelian group with the face maps $d_i:
C^{n-1}(E,C) \rightarrow C^n(E,C)$ defined for $\varphi: E \rightarrow
C^{\otimes(n-1)}$ by
\begin{equation*}
d_i(\varphi) = \begin{cases} E \stackrel{l}{\longrightarrow} C \otimes E
\stackrel{\,\,C \otimes \varphi}{\longrightarrow} C \otimes C^{\otimes n-1} =
C^{\otimes n} &(i = 0) \\
	E \stackrel{\varphi}{\longrightarrow} C^{\otimes n -1}
\stackrel{\Delta^{(i)}}{\longrightarrow} C^{\otimes n} & (0 < i < n) \\
	E \stackrel{r}{\longrightarrow} E \otimes C \stackrel{\varphi
\otimes C}{\longrightarrow} C^{\otimes n} & (i = n).
\end{cases}
\end{equation*}
Here $l$ and $r$ are the left and right coactions, and $\Delta^{(i)} =
C^{\otimes (i-1)} \otimes \Delta \otimes C^{\otimes (n-i-1)}$.  Note that this
cohomology $H^*(E,C)$ is \emph{contravariant} in $E$ and \emph{covariant}
in $C$.

In particular, given ``linear'' maps $\sigma_1, \sigma_2: C \rightarrow C$,
we can view $C$ itself as a $C$-bimodule
% the following doesn't work $\sideset{_{\sigma_1}}{_{\sigma_2}}C$, so I
% have made a hack.
\hspace{0.2cm}$C_{\sigma_2\hspace{-0.8cm}\sigma_1}$ \hspace{0.25cm}
, with left action $C
\stackrel{\Delta}{\longrightarrow} C \otimes C \stackrel{\sigma_1 \otimes
C}{\longrightarrow} C \otimes C$ and right action $C
\stackrel{\Delta}{\longrightarrow} C \otimes C \stackrel{C \otimes
\sigma_2}{\longrightarrow} C \otimes C$.  We denote the corresponding
cohomology by
\begin{equation}
\label{s4.eq1}
	HH^*_{\sigma_1,\sigma_2}(C).
\end{equation}
A map $\varphi: C \rightarrow C$ is a $1$-cocycle for this cohomology
precisely when
\begin{equation}
\label{s4.eq2}
	\Delta \circ \varphi = ( \sigma_1 \otimes \varphi + \varphi \otimes
	\sigma_2) \Delta.
\end{equation}

Now let us go back to the context of a Hopf operad $\hoP$ on our underlying
category $\C$.

\subsection{Natural twisting functions.}

Call $\sigma$ a natural twisting function if $\sigma$ assigns to each Hopf
$\hoP$-algebra $C$ a linear endomorphism $\sigma = \sigma^{(C)} : C
\rightarrow C$, which is natural for morphisms of augmented $\hoP$-algebras
(i.e. if $f: C \rightarrow D$ is such a morphism then $f \circ \sigma^{(C)}
= \sigma^{(D)} \circ f$), and has the property that $\sigma^{(k)}$ is the
identity.  Note that this implies that $\varepsilon \circ \sigma^{(C)} =
\varepsilon$.  For example, the identity $C \rightarrow C$ and the
composition $C \stackrel{\varepsilon}{\longrightarrow} k
\stackrel{u}{\longrightarrow} C$ of the augmentation and the unit are natural
twisting functions, as is any convex combination $\alpha \cdot \id + \beta
\cdot u \varepsilon : C \rightarrow C$ of these two (for $\alpha, \beta :
k \rightarrow k$ with $\alpha + \beta = \id$).

Now let $(H, \lambda)$ be the initial $\hoP[t]$-algebra, and let $\sigma_1
= \sigma_1^{(H)}, \sigma_2 = \sigma_2^{(H)} : H \rightarrow H$ be the
components of two natural twisting functions.  Suppose that $\sigma_1$ and
$ \sigma_2$ define a Hopf $\hoP$-algebra structure $(H, \Delta,
\varepsilon)$ on $H$, by Theorem 3.4.  Observe that the defining
equation $(\sigma_1, \sigma_2) \Delta = \Delta \lambda$ for the coproduct
states precisely that $\lambda$ is a $1$-cocycle for $HH^*_{\sigma_1,
\sigma_2}(H)$.  The following theorem is now a consequence of the universal
property (3.1) of $(H, \lambda)$.

\theorem{
The map $\lambda$ is the universal $1$-cocycle.  More explicitly, if $B$ is a
Hopf $\hoP$-algebra and $\gamma$ is a $1$-cocycle in the complex defining
$HH^*_{\sigma_1, \sigma_2}(B)$, there is a unique Hopf $\hoP$-algebra map
$c_{\gamma} : H \rightarrow B$ such that $c_{\gamma} \circ \lambda = \gamma
\circ c_{\gamma}$.
}

\vspace{\afterlemmaskip}

\begin{proof}
By the universal property of $H$ and $\lambda$, there is a unique
$\hoP$-algebra map $c = c_{\gamma} : H \rightarrow B$ such that $\gamma c =
c \lambda$.  It suffices to show that $c$ is a coalgebra map.  First, we
show that $c$ is a map of augmented algebras, i.e. $\varepsilon \circ c =
\varepsilon$.  By initiality of $(H, \lambda)$, it suffices to show that
the composite $(H, \lambda) \stackrel{c}{\longrightarrow} (B, \gamma)
\stackrel{\varepsilon}{\longrightarrow} (k, 0)$ is a map of
$\hoP[t]$-algebras; in other words, that $\varepsilon \gamma = 0$.  To prove
this, apply $\varepsilon \otimes \varepsilon$ to the cocycle condition
$\Delta \gamma = (\sigma_1 \otimes \gamma + \gamma \otimes \sigma_2)
\Delta$.
Using that $(\varepsilon \otimes \varepsilon) \Delta = \varepsilon$, and
$\varepsilon \sigma_i = \varepsilon$ (as observed above), this yields
$\varepsilon \gamma = (\varepsilon \otimes \varepsilon \gamma + \varepsilon
\gamma \otimes \varepsilon) \Delta = \varepsilon \gamma + \varepsilon
\gamma$.  Thus $\varepsilon \gamma = 0$, as desired.

Next, we show that the map $c$ preserves coproducts.  Observe that, by
initiality of $(H, \lambda)$, the square
\begin{center}
\mbox{
\xymatrix{
{(H, \lambda)} \ar[r]^-{\Delta} \ar[d]_c &
	{(H \otimes H, \sigma_1^{(H)} \otimes \lambda + \lambda \otimes
	\sigma_2^{(H)})} \ar[d]^{c \otimes c} \\
{(B, \gamma)} \ar[r]^-{\Delta} &
	{(B \otimes B, \sigma_1^{(B)} \otimes \gamma + \gamma \otimes
	\sigma_2^{(B)})}
}
}
\end{center}
necesarily commutes as soon as all four maps are $\hoP[t]$-algebra
homomorphisms.  The map $c \otimes c$ is the only one for which this still
has to be shown.  But, we have just proved that $c$ is a map of augmented
$\hoP$-algebras, so $c \circ \sigma_i^{(H)} = \sigma_i^{(B)} \circ c$ by
naturality.  Since alse $c \lambda = \gamma c$, the map $c \otimes c$ is
indeed a map of $\hoP[t]$-algebras.  This completes the proof of the
theorem.
\end{proof}

\section{Remarks on functoriality.}

We continue to work in the context of Hopf operads on a category $\C$ as in
\ref{ss1.1}.

\subsection{Adjoint functors.}

Let $\varphi: \hoQ \rightarrow \hoP$ be a map of Hopf operads.
Then $\varphi$ induces functors $\varphi^* : (\hoP\text{-algebras})
\rightarrow (\hoQ\text{-algebras})$ and $\overline{\varphi}^*: (\text{Hopf
} \hoP\text{-algebras}) \rightarrow (\text{Hopf } \hoQ\text{-algebras})$.
Also, $\varphi$ gives a functor $\varphi^*\!:\! (\hoP[t]\text{-algebras})
\rightarrow (\hoQ[t]\text{-algebras})$, by $\varphi^*(B, \beta) =
(\varphi^*(B), \beta)$.  If the relevant coequalizers exists in $\C$
then the first functor $\varphi^*$ has a left adjoint $\varphi_! :
(\hoQ\text{-algebras}) \rightarrow (\hoP\text{-algebras})$,
see e.g.  \cite{GJ}.  Note that $\varphi^*(k) = k$ and that
the (first) functor $\varphi^*$ commutes with tensor products
of algebras.  Hence by adjointness, there are canonical maps of
$\hoP$-algebras $\varphi_!(k) \rightarrow k$ and $\varphi_! (A
\otimes B) \rightarrow \varphi_!(A) \otimes \varphi_!(B)$.  Using
these maps, one obtains a lifting of $\varphi_!$ to a left adjoint
$\overline{\varphi}_! : (\text{Hopf-}\hoP\text{-algebras}) \rightarrow
(\text{Hopf-}\hoQ\text{-algebras})$ for $\overline{\varphi}^*$.

Now let $(H, \lambda)$ be the initial $\hoP[t]$-algebra and $(K, \mu)$ the
one for $\hoQ$.  Let $j_0: (K, \mu) \rightarrow (\varphi^*(H), \lambda)$
be the unique map of $\hoQ[t]$-algebras, and note that this is a map of
augmented $\hoQ$-algebras.  Let $j: \varphi_!(K) \rightarrow H$ be the
adjoint map; this is a map of augmented $\hoP$-algebras.  Next, consider
natural twisting functions $\sigma_1, \sigma_2$ on $\hoQ$-algebras.
These also induce $\sigma_i: H \rightarrow H$ on any $\hoP$-algebra $H$,
by $\sigma_i = \sigma_i^{(\varphi^*(H))}$.

\proposition{
Suppose $\sigma_1$ and $\sigma_2$ satisfy the conditions of Theorem
3.4 so as to make $H$ and $K$ into Hopf $\hoP$-(respectively
$\hoQ$-)algebras.  Then $j_0: K \rightarrow \varphi^*(H)$ and $j:
\varphi_!(K) \rightarrow H$ are maps of Hopf $\hoP$-(resp.
$\hoQ$-)algebras.
}

\vspace{\afterlemmaskip}

\begin{proof}
The second assertion for $j$ follows from the first for $j_0$ by
adjointness.  To see that the map $j_0$ preserves the coproduct, simply
apply initiality of $(K, \mu)$ to the square
\begin{center}
\mbox{
\xymatrix{
(K, \mu) \ar[r]^-{\Delta} \ar[d]_{j_0}  & {(K \otimes K, \sigma_1^{(K)}
\otimes \mu + \mu \otimes \sigma_2^{(K)}) } \ar[d]^{j_0 \otimes j_0} \\
{(\varphi^*(H), \lambda)} \ar[r]^-{\Delta} & {(\varphi^*(H) \otimes
\varphi^*(H), \sigma_1^{(H)} \otimes \lambda + \lambda \otimes
\sigma_2^{(H)}),}
}
}
\end{center}
exactly as in the proof of Theorem 4.2.
\end{proof}

\subsection{The operad $\oB$.}

A \emph{pointed object} is an object $X$ of $\C$ equipped with a
``basepoint'' $u: k \rightarrow X$.  We call $X$ \emph{well-pointed}
if $X$ is equipped with an augmentation $\varepsilon: X \rightarrow k$
with $\varepsilon u = \id$.  Such an object splits as $X = k \oplus
\tilde{X}$ where $\tilde{X} = \Ker(\varepsilon)$.  Let $\oB$ be the
operad whose algebras are pointed objects.  If $\hoP$ is any (Hopf)
operad then the unit of $\hoP$ gives a map of operads $u: \oB \rightarrow
\hoP$.  We consider the left adjoint $u_!$ of the induced functor $u^*:
(\hoP\text{-algebras}) \rightarrow (\oB\text{-algebras})$.

\lemma{
If $X$ is well-pointed then $u_!(X) = F_{\hoP}(\tilde{X})$, the free
$\hoP$-algebra on $\tilde{X}$.
}

\vspace{\afterlemmaskip}

\begin{proof}
Let $k \stackrel{u}{\longrightarrow} X
\stackrel{\varepsilon}{\longrightarrow} k$ be a well-pointed object.
Let $w: X \rightarrow F_{\hoP}(\tilde{X}) = F(\tilde{X})$ be the
map $k \oplus \tilde{X} \rightarrow F(\tilde{X})$ obtained from the
unit $u_{F(\tilde{X})} : k \rightarrow F(\tilde{X})$ of this free
algebra together with the canonical map $\mu : \tilde{X} \rightarrow
F(\tilde{X})$.  We claim that $w$ is the universal base-point preserving
map from $X$ into a $\hoP$-algebra.  Indeed, suppose $f: X \rightarrow
A$ is any map into the underlying object $A$ of a $\hoP$-algebra
$\underline{A}$, with $f \circ u = u_{\underline{A}}$.  Since
$F(\tilde{X})$ is the free algebra, the restriction $f \upharpoonright
\tilde{X} : \tilde{X} \rightarrow A$ extends uniquely to a $\hoP$-algebra
map $\underline{f} : F(\tilde{X}) \rightarrow \underline{A}$.  It is easy
to check that $\underline{f} \circ w = f$ for this map $\underline{f}$.
\end{proof}

Now let $(A, \alpha)$ be the initial $\oB[t]$-algebra, and $(H, \lambda)$
the initial $\hoP[t]$-algebra as before.  Let $\sigma_1, \sigma_2$ be
natural twisting functions on $\oB$-algebras.  Suppose $\sigma_1^{(A)},
\sigma_2^{(A)}: A \rightarrow A$ define a Hopf algebra structure on $A$,
and $\sigma_1^{(H)}, \sigma_2^{(H)} : H \rightarrow H$ one on $H$, by
Theorem 3.4.

\proposition{
There is a canonical retraction
\begin{center}
\mbox{
\xymatrix{
	{u_!(A)} \ar@<1mm>[r]^-{j} & {H, \quad r \circ j = \id,}
	\ar@<1mm>[l]^-{r}
}
}
\end{center}
where $j$ is a map of Hopf $\hoP$-algebras and $r$ one of augmented
$\hoP$-algebras.
}

\vspace{\afterlemmaskip}

\begin{proof}
The map $j: u_!(A) \rightarrow H$ is the one of Proposition 5.2.
The map $r : H \rightarrow u_!(A)$ is the unique map $(H, \lambda)
\rightarrow (u_!(A), \overline{\alpha})$ of $\hoP[t]$-algebras, for the
map $\overline{\alpha}$ defined as follows.  Since $A$ has an augmentation
$\varepsilon$ with $\varepsilon \alpha = 0$ (Lemma 3.2), we can write
$A = k \oplus \tilde{A}$ where $\alpha$ maps $A$ into $\tilde{A}$.
Also, the free $\hoP$-algebra $u_!(A) = F_{\hoP}(\tilde{A})$, briefly
$F(\tilde{A})$, is augmented, hence splits as $u_!(A) = k \oplus
F(\tilde{A})\! \tilde{\phantom{A}}$.  Now define $\overline{\alpha}$
on these two summands separately: on $k$ it is the composition
\begin{equation*}
	k \stackrel{u}{\longrightarrow} A \stackrel{\alpha}{\longrightarrow}
	\tilde{A} \rightarrow F(\tilde{A})
\end{equation*}
and on the other summand it is the map
\begin{equation*}
	F(\tilde{A})\! \tilde{\phantom{A}} \subseteq F(\tilde{A})
	\stackrel{F(\tilde{\alpha})}{\longrightarrow} F(\tilde{A})
\end{equation*}
where $\tilde{\alpha} : \tilde{A} \rightarrow \tilde{A}$ is the
restriction of $\alpha$.  Note that the map $\overline{\alpha}$ thus
defined satisfies the identities
\begin{equation*}
	\overline{\alpha} w = w \alpha, \ \varepsilon \overline{\alpha} = 0,
\end{equation*}
where $w: A \rightarrow u_!(A)$ is the universal map as in the proof
of the previous lemma.

We claim that $r \circ j = \id$.  By adjointness, it suffices to show
$rjw=w$ as maps of pointed objects.  Now $w \alpha = \overline{\alpha}
w$ as we have seen.  Also, $j: u_!(A) \rightarrow H$ is obtained
from $j_0 : A \rightarrow u^*(H)$ by adjointness, hence $j w = j_0$.
Thus $(rjw) \alpha = rj_0 \alpha = r \lambda j_0 = \overline{\alpha}
r j_0 = \overline{\alpha}(rjw)$.  This shows that $w$ and $rjw$ are both
maps of $\oB[t]$-algebras on $(A, \alpha)$, hence equal by initiality.

It remains to observe that $r$ respects the augmentation.  Since $r :
(H, \lambda) \rightarrow (u_!(A), \overline{\alpha})$ and $\varepsilon
: (u_!(A), \overline{\alpha}) \rightarrow (k, 0)$ are both maps of
$\hoP[t]$-algebras, so is the composite $\varepsilon r$.  So $\varepsilon
r = \varepsilon$ by initiality of $(H, \lambda)$.  This shows that $r$
preserves the augmentation, and completes the proof.
\end{proof}

\subsection{Example.}

Let $(H, \lambda)$ be the Connes-Kreimer Hopf algebra of Example
3.6. 
For the same twisting functions $\sigma_1 = \id$ and $\sigma_2 =
u \varepsilon$,
the initial $\oB[t]$-algebra $(A, \alpha)$ is the vector
space with basis $x_0, x_1, x_2, \ldots$, where $x_0$ is the base point and
$\alpha(x_n) = x_{n+1}$.  Thus $u_!(A)$ is the algebra $k[x_1, x_2, \ldots]$,
where we identify $x_0$ with $1 \in u_!(A)$.  The Hopf algebra structure is
given by $\Delta(x_n) = \sum_{i=0}^{n} x_i \otimes x_{n-i}$.  The embedding
$j$ identifies $u_!(A)$ with the subalgebra of ``linear trees'' of $H$
(considered also in \cite{CK}), and
$x_n$ with $\lambda^n(1) \in H$.  The retraction $r : H \rightarrow u_!(A)$
sends a tree $T$ to the product of all the maximal branches through $T$.
For example, the tree
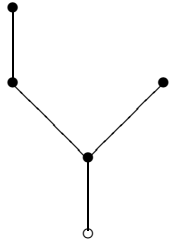
\[
\begin{xy}
%points
<0cm,3cm>="LU"*{\bullet},
<0cm,2cm>="L"*{\bullet},           <2cm,2cm>="R"*{\bullet},
             <1cm,1cm>="C"*{\bullet},
             <1cm,0cm>="B"*{\circ},
%lines
\ar@{-} "LU";"L",
\ar@{-} "L";"C",    \ar@{-} "R";"C",
         \ar@{-} "C"; "B"+<0cm,0.5mm>
\end{xy}
\]
representing $\lambda(\lambda^2(1) \cdot \lambda(1))$ is sent to $x_3 \cdot
x_1$.  Note that $r$ does not commute with coproducts.

\vspace{0.5cm}

\noindent Utrecht, April 1999.

\end{document}